\documentclass{webofc}
\usepackage[varg]{txfonts}   

\usepackage{epsfig,amsmath,amssymb,slashed}
\usepackage[utf8]{inputenc}
\usepackage{hyperref}
\usepackage{xcolor}
\usepackage{bm}
\usepackage{combelow}
\usepackage[normalem]{ulem}

\newcommand{\wad}[2]{\widehat a^\dagger_{#2}(#1)}
\newcommand{\wa}[2]{\widehat a_{#2}(#1)}
\newcommand{\di}{{\rm d}}
\newcommand{\Tr}{{\rm Tr}}

\def\wQ{{\widehat Q}}
\def\wP{{\widehat P}}
\def\wJ{{\widehat J}}
\def\wrho{{\widehat{\rho}}}
\newcommand{\tr}{{\rm tr}}  
\newcommand{\e}{{\rm e}}

\begin{document}
\title{Exact Polarization of Particles of Any Spin at Global Equilibrium}

\author{\firstname{Andrea} \lastname{Palermo}\inst{1}\fnsep\thanks{\email{andrea.palermo@stonybrook.edu}}} 

\institute{Center for Nuclear Theory, Department of Physics and Astronomy,
Stony Brook University, Stony Brook, New York 11794–3800, USA }

\abstract{%
  The polarization of the $\Lambda$ particle offers the unique opportunity to study the hydrodynamic gradients in the Quark-Gluon Plasma formed in heavy-ion collisions. However, the theoretical formula commonly used to calculate polarization is only a linear order expansion in thermal vorticity and neglects higher-order corrections. Here, I present an exact calculation to all orders in (constant) thermal vorticity at global equilibrium, obtaining the analytic form of the spin density matrix and the polarization vector for massive particles of any spin. Finally, I extend these results to local equilibrium and assess their phenomenological impact by numerically calculating the polarization vector in a 3+1 hydrodynamic simulation.
}
\maketitle
\section{Introduction}
The experimental measurement of the polarization vector of $\Lambda$ particles in heavy ion collisions paved a new way in the hydrodynamic studies of the QGP \cite{STAR:2017ckg,STAR:2019erd,STAR:2020xbm,ALICE:2021pzu,ALICE:2019onw,Florkowski:2017ruc,Florkowski:2018fap,Weickgenannt:2019dks,Weickgenannt:2020aaf,Gao:2020lxh,Becattini:2020ngo,Becattini:2022zvf,Gao:2020vbh,Hongo:2021ona}. In fact, with spin polarization one can access the thermodynamic gradients in the plasma at freeze-out. According to the hydrodynamic model and based on a statistical calculation, it was found early on that the polarization vector of massive spin-S fields is \cite{Becattini:2013fla,Fang:2016vpj,Becattini:2019ntv}:
\begin{equation}\label{eq:polarization linear}
    S^\mu(p)=-\frac{S(S+1)}{3}\frac{1}{2m}\epsilon^{\mu\nu\rho\sigma}p_\sigma\frac{\int\di\Sigma\cdot p 
    \,\varpi_{\nu\rho}n(p)[1+(-1)^{2S} n(p)]}{\int\di\Sigma\cdot p \,n(p)},
\end{equation}
where $p$ is the four-momentum of the particle at hand, the integral is calculated on the decoupling hypersurface, $n(p)=[\e^{\beta\cdot p}+(-1)^{2S+1}]^{-1}$ is the Fermi-Dirac or the Bose-Einstein distribution function and $\varpi$ is the thermal vorticity:
\begin{equation}
    \varpi_{\mu\nu}=-\frac{1}{2}(\partial_\mu\beta_\nu-\partial_\nu\beta_\mu),
\end{equation}
$\beta^\mu$ being the four-temperature vector.  

Additional contributions to spin polarization have been found later on, including the thermal shear \cite{Becattini:2021suc,Becattini:2021iol,Fu:2021pok,Liu:2021uhn} and the spin hall effect \cite{Fu:2022myl,Ivanov:2020udj,Buzzegoli:2022qrr}, which I will not discuss further here. 
It is important to stress that all the terms mentioned so far are the leading order contributions to the spin vector in linear response theory. Even in the case of global equilibrium, where the only contribution to the spin vector comes from the constant thermal vorticity, there are no results beyond the first order of approximation in $\varpi$. Here, I show how to compute the spin vector to all orders in $\varpi$ for fields of any spin at global equilibrium \cite{Palermo:2023cup}. I will also provide a finite and analytic expression of the spin density matrix, which can be used to compute any spin observable, including the vector meson spin alignment.

\section{Spin density matrix and polarization vector}
\label{sec-1}
To compute the polarization vector or any other spin-related quantity, one uses the spin density matrix $\Theta$. For massive fields of spin $S$, the spin density matrix is a $(2S+1)\times(2S+1)$ hermitian matrix, characterized by $S^2-1$ real numbers. The definition of the spin density matrix adopted in the context of relativistic heavy-ion collisions is \cite{becattini2020polarization}:
\begin{align}\label{definition theta field theory}
    \Theta_{sr}(p)=\frac{\Tr\left(\wrho\,\wad{p}{r}\wa{p}{s}\right)}
    {\sum_{l}\Tr\left(\wrho\,\wad{p}{l}\wa{p}{l}\right)}=
    \frac{\langle\wad{p}{r}\wa{p}{s}\rangle}{\sum_l\langle\wad{p}{l}\wa{p}{l}\rangle}.
\end{align}
In the above equation, $\wrho$ is the statistical operator, and $\widehat{a}$ and $\widehat{a}^\dagger$ are the annihilation and creation operators of states with momentum $p$ and spin $r$, $s$. 

The spin density matrix is related to the spin vector $S^\mu$ through \cite{becattini2020polarization}:
\begin{align}\label{eq:spin vect tr J}
S^\mu(p)=\sum_{i=1}^3 [p]^\mu_{\ i}\tr\left(\Theta(p) D^S(\mathrm{J}^i)\right),
\end{align}
where $D^S(\mathrm{J}^i)$ is the $i$-th generator of $\text{SO}(3)$ in the $S$-representation. The transformation denoted as $[p]$ is the so-called \emph{standard boost}: a Lorentz transformation that maps the rest frame momentum $\mathfrak{p}=(m,\bm{0})$ to the momentum $p=(\varepsilon,\bm{p})$, i.e. $p^\mu = [p]^\mu_{\ \nu}\mathfrak{p}^\nu$. Other formulae for the polarization vector, such as those involving the Wigner function, are always derived from Eqs. \eqref{definition theta field theory} and \eqref{eq:spin vect tr J}. 

\section{Exact spin physics at global equilibrium}
From equation \eqref{eq:spin vect tr J}, the polarization vector can be computed once the spin density matrix is known. Here, I will compute exactly the spin density matrix for fields of any spin at global equilibrium.

I will consider the following density operator:
\begin{equation}\label{eq:density op glob eq}
    \wrho=\frac{1}{Z}\exp\left[-b\cdot \wP+\frac{\varpi:\wJ}{2}+\zeta \wQ\right],
\end{equation}
where $\wQ$ is the charge operator, and $\wP$ and $\wJ$ are the generators of the Poincaré group: the four-momentum and the angular momentum-boost operator, respectively. The vector $b^\mu$ and antisymmetric tensor $\varpi^{\mu\nu}$, that is the thermal vorticity, are constants, and they define the four-temperature vector as the Killing vector:
\begin{align*}
    \beta^\mu(x)=b^\mu + \varpi^{\mu\nu}x_\nu.
\end{align*}
In addition, $\zeta=\mu/T$ is also constant, $\mu$ being the chemical potential. The proper temperature is $T=\left(\sqrt{\beta\cdot \beta}\right)^{-1}$. 

In Refs.~\cite{Becattini:2020qol,Palermo:2021hlf}, it was realized that factorizing the operator \eqref{eq:density op glob eq} and performing the analytic continuation $\varpi\mapsto -i\phi$ provides a powerful technique to compute exact expectation values. This method allows the calculation of $\langle\wad{p}{r}\wa{p}{s}\rangle$:
\begin{equation}\label{eq: exact number operator}
    \langle \wad{p}{s} \wa{p'}{t}\rangle=2\varepsilon' \, \displaystyle \sum_{n=1}^{\infty}(-1)^{2S(n+1)}
    \delta^3(\Lambda^n{\bf{p}}-{\bf{p}'})D(W(\Lambda^n,p))_{ts}
    \e^{-\widetilde{b}\cdot\sum_{k=1}^n\Lambda^kp}\e^{n\zeta},
\end{equation}
 where $\Lambda=\exp\left[-i\phi:J/2\right]$ is the Lorentz transformation with parameter $\phi$ equal to the \emph{imaginary} vorticity and $W(\Lambda,p)=[\Lambda p]^{-1}\Lambda [p]$ is the so-called Wigner rotation. The vector $\tilde{b}(\phi)$ is:
\begin{equation}\label{eq: def b tilde}
    \tilde{b}^{\mu}(\phi)=\sum_{k=0}^{\infty}\frac{1}{(k+1)!}\phi^{\mu}_{\ \nu_1}
    \phi^{\nu_1}_{\ \nu_2}\dots\phi^{\nu_{k-1}}_{\ \nu_k}b^{\nu_k},
\end{equation}
and its form is a consequence of the fact that the Poincaré group is non-Abelian. 

Plugging eq.~\eqref{eq: exact number operator} into eq.~\eqref{definition theta field theory}, yields:
\begin{equation}
    \Theta_{rs}(p)
=\frac{\sum_{n=1}^\infty (-1)^{2S(n+1)} 
\e^{-\widetilde{b}\cdot\sum_{k=1}^n\Lambda^kp}\e^{n\zeta}D_{rs}(W(\Lambda^n,p))\delta^3(\Lambda^np-p)}
{\sum_{n=1}^\infty (-1)^{2S(n+1)} \e^{-\widetilde{b}\cdot
\sum_{k=1}^n\Lambda^kp}\e^{n\zeta}\tr[D(W(\Lambda^n,p))]\delta^3(\Lambda^np-p)}.
\end{equation}
To proceed further in the calculation, and as long as the thermal vorticity is imaginary, the constraint $\Lambda p = p$ must be enforced. Indeed, from eq.~\eqref{eq: exact number operator}, one sees that if this is not the case $\langle \wad{p}{s} \wa{p}{t}\rangle=0$, and the spin density matrix would become trivial. 

Imposing the constraint $\Lambda p =p$, it follows that:
\begin{align}
    \phi^{\mu\nu} = \epsilon^{\mu\nu\rho\sigma}\xi_\rho \frac{p_\sigma}{m}, && \xi^\rho=-\frac{1}{2m}\epsilon^{\rho\mu\nu\sigma}\phi_{\mu\nu}p_\sigma.
\end{align}
Thanks to this simplification, and after some algebra (see Ref.~\cite{Palermo:2023cup} for the full derivation), the spin density matrix becomes:
\begin{align*}
\Theta(p)=\frac{\sum_{n=1}^\infty (-1)^{2S(n+1)} \e^{-nb \cdot p}\e^{n\zeta}\e^{-i n\bm{\xi}_0\cdot D^S(\textbf{J})}}
{\sum_{n=1}^\infty (-1)^{2S(n+1)} \e^{-nb\cdot p}\e^{n\zeta}\tr\left(\e^{-i n\bm{\xi}_0\cdot D^S(\textbf{J})}\right)},
\end{align*}
where $\xi_0^\mu={[p]^{-1}}^{\mu}_{\ \nu}\xi^\nu$, $\bm{\xi}_0$ being its spatial part, and $D^S(\textbf{J})$ is the three-vector of the generators of the rotation group in the $S$-representation. As long as the absolute value of the eigenvalues of the matrix $\exp(-b\cdot p+\zeta)\exp\left[-i\bm{\xi}_0\cdot D^S(\textbf{J})\right]$ is less than one (that is $b\cdot p>\zeta$) the series can be summed as a geometric series of matrices \cite{hubbard2002vector}. Defining:
\begin{equation}\label{eq: definition theta vector}
    \theta^\mu =-\frac{1}{2m}\epsilon^{\mu\nu\rho\sigma}\varpi_{\nu\rho}p_\sigma,
\end{equation} 
the spin density matrix after the analytic continuation $\phi\mapsto i\varpi$ reads:
\begin{equation}\label{eq: exact theta}
    \Theta(p)=\frac{ \left[(-1)^{2S}\mathbb{I}- \e^{-b\cdot p+\bm{\theta_0} \cdot  D^S(\textbf{J})}\right]^{-1}-(-1)^{2S}\mathbb{I}}
{\sum_{k=-S}^S \left(\e^{b\cdot p-k \sqrt{-\theta^2}}-(-1)^{2S}\right)^{-1}}.
\end{equation}

From eq.~\eqref{eq: exact theta}, one can compute all spin-related quantities, including spin polarization and spin alignment. In particular, the mean spin vector reads, using eq.~\eqref{eq:spin vect tr J}:
\begin{equation}\label{eq: spin sum of statistics}
    S^\mu(p)=\frac{\theta^\mu}{\sqrt{-\theta^2}}\frac{\sum_{k=-S}^{S} k  
    \left[ \e^{b\cdot p-\zeta - k\sqrt{-\theta^2}}-(-1)^{2S} \right]^{-1}}
    {\sum_{k=-S}^{S} \left[ \e^{b\cdot p-\zeta - k\sqrt{-\theta^2}}-(-1)^{2S} \right]^{-1}}.
\end{equation}
This formula reproduces the previous literature, including the small vorticity approximation and the case of Boltzmann statistics. In the case of very large thermal vorticity, it predicts a polarization equal to one \cite{Palermo:2023cup}. 

\section{Polarization in heavy-ion collisions}
\label{sec-2}
One can now evaluate the effect of the formula \eqref{eq: spin sum of statistics} in phenomenological applications. Although the derivation holds in global equilibrium, eq.~\eqref{eq: spin sum of statistics} can be generalized to local equilibrium by promoting the thermal vorticity to be a local variable, i.e. $\varpi\mapsto \varpi(x)$, and integrating eq.~\eqref{eq: spin sum of statistics} over the decoupling hypersurface.
\begin{equation}\label{eq:local polarization}
S^\mu_{LE}(p)=\frac{\int\di\Sigma\cdot p\,
n(p)\, S^\mu(p)}{\int\di\Sigma\cdot p\,n(p)}.
\end{equation}

\begin{figure}
    \centering
    \includegraphics[width=0.45\textwidth]{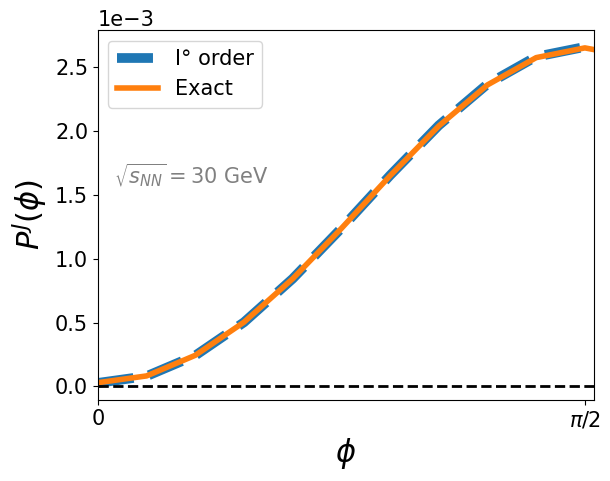}  \includegraphics[width=0.45\textwidth]{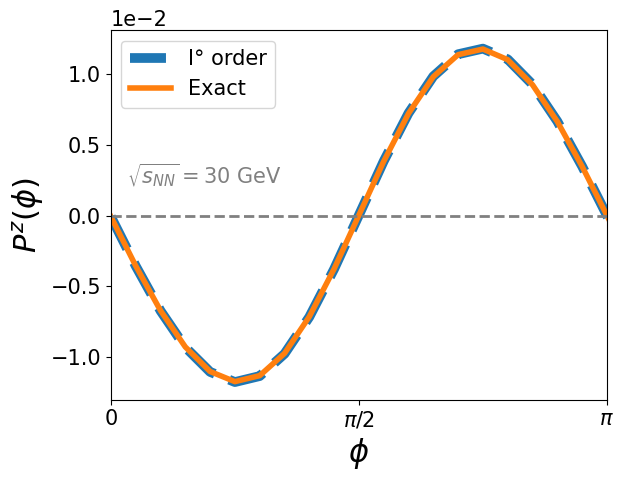}
    \caption{The components of the polarization vector along the total angular momentum (left panel) and the beam direction (right panel) as a function of the azimuthal angle. The blue dashed line is the linear approximation \eqref{eq:polarization linear}, whereas the solid orange line corresponds to Eq.~\eqref{eq:local polarization}.}
\label{fig:pol}
\end{figure}
Here, in Fig.~\ref{fig:pol}, I report the result of Eq.~\eqref{eq:local polarization} compared to that of Eq.~\eqref{eq:polarization linear} in Au-Au collisions at $\sqrt{s_{NN}}=30$ GeV in the centrality class 10-60\%, simulated using the code vHLLE \cite{Karpenko:2013wva} for the hydrodynamic evolution. The initial state is an averaged entropy density profile generated by GLISSANDO v.2.702 \cite{Rybczynski:2013yba}. For the figure, I consider the case of Dirac fermions at midrapidity.
I remind the reader that the polarization is related to the spin vector by $P^\mu =S^\mu/S$, where $S$ is the spin. One can see that the two formulae give the same result, and it follows that the linear approximation is a good one, at least for $\sqrt{s_{NN}}>30$ GeV.

\section{Conclusions}
In this work, I have shown how to calculate the spin density matrix and the polarization vector for massive spin-S fields to all orders in thermal vorticity and obtained an analytic finite expression for both quantities. Using the spin density matrix, all spin observables can be computed, including alignment. The polarization vector is a finite sum, representing the average of the spin state weighted by Bose or Fermi distribution functions.

I have shown a numerical calculation assessing the importance of high-order corrections in vorticity to the phenomenology of heavy ion collisions. As it turns out, the exact solution and the linear approximation precisely coincide for AuAu collisions at $30$ GeV. Yet, a small discrepancy may arise in lower-energy collisions, where the thermal vorticity is larger. This work shows that the linear approximation of the polarization vector is an excellent one in most practical applications to heavy-ion collisions.

\bibliography{biblio}
\end{document}